\documentclass[article,floats,prl,aps]{revtex4}

\begin{document}

\preprint{\hbox { } }

\title{Primordial Black Holes, Hawking Radiation and the Early Universe}
\author{\bf Paul H. Frampton$^{(a)}$ and Thomas W. Kephart$^{(b)}$}
\address{(a)Department of Physics and Astronomy,\\
University of North Carolina, Chapel Hill, NC  27599.}
\address{(b)Department of Physics and Astronomy,\\
Vanderbilt University, Nashville, TN 37325.}

\bigskip
\bigskip
\bigskip
\bigskip
\bigskip

\date{\today}

\bigskip
\bigskip
\bigskip
\bigskip

\begin{abstract}
The 511 keV gamma emission from the galactic
core may originate from a high concentration ($\sim 10^{22}$) of
primordial black holes (PBHs) in the core each of whose Hawking radiation
includes $\sim 10^{21}$ positrons per second. The PBHs we consider
are taken as near the lightest
with longevity greater than the age of the universe
(mass $\sim 10^{12}$ kg; Schwarzschild radius $\sim 1$ fm).
These PBHs contribute only a small fraction of cold dark matter,
$\Omega_{PBH} \sim 10^{-8}$. This speculative hypothesis, if confirmed implies the simultaneous discovery of Hawking radiation and an early universe phase transition.
\end{abstract}

\maketitle

\newpage

\bigskip
\bigskip


\bigskip

The slices of the cosmological energy pie have been
clarified recently. Only about 4\% comes from the
familiar baryons. The other 96\% is mysterious:
some 25\% is dark matter (hereafter DM) and the remaining 71\% is
dark energy. Neither of these two large slices
is understood.

\bigskip

Here we investigate the hypothesis that
a small fraction of cold dark matter (CDM) is made up of
primordial black holes (PBHs). In particular, we investigate
the possibility that the 511 KeV gamma rays from the vicinity of Sagittarius A*
may originate from Hawking radiation\cite{Hawking}; the
detection of Hawking radiation within the Solar System
is frustrated by the number (as we shall estimate below)
of PBHs therein.

We should first define
the meaning of PBH.
The Hawking temperature of a Black Hole is given by
\begin{equation}
T_H = \frac{M_{Planck}^2}{8 \pi M} \sim 4 \times 10^{-10} \left( \frac{M_{Planck}}{M} \right) kg
\label{TH}
\end{equation}
and, using $1 kg \sim 10^{27} eV$, $1^oK \sim 10^{-4} eV$ this is
\begin{equation}
T_H \sim 10^{31}K \left( \frac{M_{Planck}}{M} \right)
\label{TH2}
\end{equation}
so to make $T_H \sim 2.7K$, the temperature of the Cosmic Background Radiation (CMB),
one needs $M \sim 3 \times 10^{30}M_{Planck} \sim 3 \times 10^{22} kg \sim (1\%) M_{\oplus}$,
or about one per cent of the mass of the Earth, $M_{\oplus} \sim 4 \times 10^{24} kg$.
Let us denote this PBH mass by $M_{rad}$ since it is the maximum mass PBH which will
radiate with a temperature higher than that of the CMB.

\bigskip

Heavier black holes will generally accrete matter
and grow but these are also possible PBHs.
In the galactic halo the maximum mass compatible
with the behavior observed \cite{XO,AMS}
we shall take
to be $M_{max} \sim 10^6 M_{\odot} \sim 10^{36} kg$.

\bigskip

The lifetime\cite{DK} of a PBH is proportional to $M^3$. For $T_H > 100GeV$
it is approximately
\begin{equation}
\tau (M) \sim 100 \tau_{Planck} \left( \frac{M}{M_{Planck}} \right)^3
\label{tau}
\end{equation}
Taking $\tau_{Planck} \sim 10^{-43} sec$ we find $\tau(M_{min}) \sim t_O
\sim 10^{18} sec$, the age of the Universe, for $M_{min} \sim 5 \times 10^{11}
kg$. This is therefore the minimal mass PBH which could have survived since
the Big Bang.

\bigskip

Thus the allowed range of masses of PBHs in existence today spans 24 order of magnitude from
$M_{max} \sim 10^{36}$ kg down to $M_{min} \sim 10^{12}$ kg.
We shall assume a phase transition in the early
universe which produces PBHs centered around $M_{PBH} = 10^{12}$ kg.

\bigskip

Black hole sizes are characterized by the Schwarzschild radius $r = 2 G M$.
It is of order $10^7$
km for $M_{max}$ and $10^{-13} cm = 1 fm$ for $M_{min}$. The minimal
PBH is a formidable object with the mass of Mount Everest
and the size of a proton; its density is some 53 orders of magnitude
times that of water, and its Hawking temperature
is a hundred billion ($10^{11}$) degrees Kelvin, or $\sim 10 $ MeV.
Such $(PBH)_{min}$ are the most interesting cases
since they may make their presence known by an explosive
final evaporation. Detection of Hawking radiation is most
likely if all PBHs have mass $M_{min}$, so we shall
emphasize this unjustified (but interesting) assumption.

\bigskip

Let us make a preliminary discussion of the CDM.
A convenient volume is $V = (mly)^3$, roughly the volume of the Solar System
where $mly$ is a milli-lightyear $\sim 10^{15}cm$. The total DM mass
in the visible Universe is
$\sim 10^{80} GeV \sim 10^{53} kg$ in a total volume $\sim 10^{39} (mly)^3$ so the
average DM density is $ \sim 10^{14} kg/(mly)^3$.
Since the DM is concentrated in clusters with an overdensity
of about a hundred, the local DM density is closer to
$10^{16} kg/ (mly)^3$ or $10^4 (PBH)_{min} / (mly)^3$.

\bigskip

We would need to estimate the mass distribution of PBHs, but
let us continue with the most optimistic scenario.
Consider the core of the Milky Way, which
subtends an angle of several degrees at a distance about $10^4 ly$
and so the core volume is $\sim (10^3 ly)^3 = 10^{18}(mly)^3$
or $\sim 10^{18}$ times the volume of the Solar System.
The galactic core thus contains some $10^{34}$ kg of DM which
corresponds to $\sim 10^{22} M_{min}$. This corresponds
to only a small fraction of the CDM, $\Omega_{CDM} \sim 10^{-8}$.

Given that the local density of CDM is $\sim 10^{16} kg/(mly)^3$,
detection of Hawking radiation within the Solar System is quickly
seen to be impracticable since our $\Omega_{PBH} \sim 10^{-8}$
suggests the number of PBHs in the Solar System is vanishingly
small (a number $\sim 10^{-4}$ !) even for the minimal
mass of a PBH, becoming even smaller if larger
$M_{PBH}$ are considered: unfortunately, confirmation of
Hawking radiation within the $mly$ scale of the Solar System looks
impossible even in this most optimistic scenario.
Therefore, we must look further afield, at least
$\sim 10^7$ times further, to
the distance ($\sim 10 kpc$) of the galactic
core where the situation is much more hopeful.

\bigskip
\bigskip

Given our hypothesis that $\Omega_{PBH} \sim 10^{-8}$
in the form of PBHs, we
have seen that there can be up to $\sim 10^{22} (PBH)_{min}$
in the core of the Milky Way. We now investigate whether
this is a plausible source of the 511 KeV gamma rays
detected therefrom, see the recent measurements
reported in \cite{SPI1,SPI2} and the earlier
data cited therein.

The mass decay rate of a PBH is given by (see {\it e.g.}
\cite{Carr:2003bj, Carr:1998fw})

\begin{equation}
\frac{dM}{dt} = - 5 \times 10^{25} (M~~~ {\rm in ~~~ grams})^{-2} g/s
\label{mdot}
\end{equation}
so for $M = 10^{15}$ g the rate of mass loss is $5 \times 10^{21}$ g/s,
which using 1 g $\sim 10^{24}$ eV and the fact that some 22\% of
the radiation\cite{Carr:2003bj} is in the form of positrons suggests some $10^{21}$
positrons per second per PBH.

Thus the total number of annihilations in the core per second is
estimated as $\sim 10^{22+21} = 10^{43}$, very consistent
\cite{footnote}
with the observational number \cite{SPI1,SPI2} which
is $1.3 \times 10^{43}$ per second.

This then provides an attractive alternative to
the ideas \cite{silk,Hooper1,Hooper2} of dark matter annihilation or \cite{Parizot:2004ph}
gamma-ray bursters as an explanation of the 511 KeV gamma rays.

\bigskip
\bigskip

Let us consider the implications of PBHs with mass $\sim 10^{12}$ kg.
In order for them to form there must have been density perturbations
of mass of order the particle horizon $M_H(t)$ mass at their time of formation, where
$M_H(t) \approx {c^3 t\over G} \approx 10^{15}\left({t\over 10^{-
23} \ {\rm s}}\right) g.$
At the Planck time $\sim 10^{-43}$ s one finds $M_H(t) \approx 10^{-5}$ g. When
$M_H(t) \approx 10^{12}$ kg we have $t\sim 10^{-20}$ s, the horizon size
was $\sim 1$ fm, and the energy scale was $\sim 10^8$ GeV.
This provides information on the state of the Universe at this
early epoch. In fact, to generate the  density perturbations the
Universe would need to have been going through a phase transition
at this time. An earlier phase transition would have generated less
massive PBHs which would have already evaporated by now. A later
phase transition would have generated more massive PBHs with $T_H$
below the level needed to evaporate positrons. So, given the evidence
for positrons from  the 511 KeV gamma ray data, and assuming the positrons
come from PBHs, we are lead to the inevitable conclusion that there must
have been a phase transition in  the early Universe at approximately
$\sim 10^8$ GeV. Furthermore, there could have been very little inflation
since this  phase transition, otherwise the PBHs would have been inflated away.
The $\sim 10^8$ GeV energy scale in near the geometric mean of the
electroweak scale and the grand unification or string scale, and has
appeared in many contexts in particle physics model building, ranging from
axion physics (which has been suggested as a dark matter candidate) to
the scale of supersymmetry breaking. PBHs at mass $\sim 10^{12}$ kg is
solid evidence for a phase transition at the $\sim 10^8$ GeV energy scale.

To confirm that the the 511 keV gamma rays are coming from PBHs
further tests are required. For example, along with the positrons,
there are electrons and a black body spectrum of direct gamma rays.
If these spectra could be discovered, then the PBH scenario would be
on solid ground and simultaneously imply the discovery of Hawking
radiation and an early universe phase transition.\\

\bigskip

\noindent {\it Acknowledgements}

\bigskip

We acknowledge communications from Professor Jane McGibbon.
This work was
supported in part by the US Department of Energy under Grants No.
DE-FG02-97ER-41036 (PHF) and No. DE-FG05-85ER40226 (TWK).


\begin{thebibliography}{99}
\bibitem{footnote}
This general idea seems to have been suggested, among a variety of
possible sources, by the discoverers\cite{Leventhal} of the 511 keV line.
\end{thebibliography}
\end{document}